\documentclass[prl, twocolumn, nofootinbib, amsmath, amssymb, aps, floatfix]{revtex4-2}
\pdfoutput=1

\usepackage{hyperref}
\usepackage{cleveref}

\usepackage[T1]{fontenc}
\usepackage[utf8]{inputenc}
\usepackage{newtxtext}
\usepackage{newtxmath}
\usepackage{microtype}

\usepackage{graphicx}
\usepackage{caption}
\usepackage{subcaption}

\usepackage[italicdiff]{physics}

\hypersetup{
    colorlinks=true,
    allcolors=[rgb]{0.76,0.21,0.18},
}

\newcommand{\given}{\,|\,}
\newcommand{\RGI}{\mathcal{I}}
\newcommand{\LQCD}{\Lambda_\text{QCD}}
\newcommand{\jacobian}{|\mathcal{J}|}

\begin{document}

\title{Origins of parameters in adimensional models}
\author{Andrew Fowlie\bigskip}
\affiliation{Department of Physics, School of Mathematics and Physics, Xi'an Jiaotong-Liverpool University, 111 Ren'ai Road, Suzhou Dushu Lake, Science and Education Innovation District, Suzhou Industrial Park, Suzhou 215123, P.R.~China}
\email{andrew.fowlie@xjtlu.edu.cn}

\begin{abstract}
We explore the origins of parameters in adimensional theories --- fundamental theories with no classical massive scales. If the parameters originate as draws from a distribution, it should be possible to write a distribution for them that doesn't depend on or introduce any massive scales. These distributions are the invariant distributions for the renormalization group (RG). If there exist RG invariant combinations of parameters, the RG invariant distributions are specified up to arbitrary functions of the RG invariants. If such distributions can be constructed, adimensional theories could predict the values of their parameters through distributions that are constrained by the RG. If they can't be constructed, the parameters must originate in a different way. We demonstrate the RG invariant distributions in QCD, the Coleman-Weinberg model and a totally asymptotically free theory.
\end{abstract}

\maketitle
\newpage

\section{Introduction}

The weak scale in the Standard Model (SM) suffers from a fine-tuning problem; if there are massive scales in nature, barring fine-tuning, any hierarchy between the weak scale and the massive scale would be destroyed by quantum corrections. To avoid this, we could assume that nature is adimensional, that is, that it contains no fundamental massive scales at all~\cite{Foot:2007iy,Heikinheimo:2013fta,Gabrielli:2013hma,Englert:2013gz,Kannike:2014mia}. In this radical approach, the known massive scales in nature are generated anomalously by dimensional transmutation~\cite{Salvio:2020axm} as a result of renormalization group (RG) running~\cite{Wilson:1971dc,Wilson:1971vs,Wilson:1973jj,Wilson:1974mb,Carvalho:2022hca}. There can be no quadratic corrections to the weak scale on dimensional grounds alone~\cite{Bardeen:1995kv}. Furthermore, if this principle is applied to gravity~\cite{Salvio:2014soa,Salvio:2017qkx} it leads to an $R^2$ theory that is renormalizable~\cite{Salvio:2018crh,Donoghue:2021cza} and could address shortcomings of the Standard Model, including inflation~\cite{Kannike:2015apa}. Although it though suffers from ghosts, quadratic gravity may be viable~\cite{Salvio:2015gsi,Raidal:2016wop,Strumia:2017dvt,Gross:2020tph,Donoghue:2021eto}.

We consider the origins of an adimensional model's parameters and, by requiring the origins to be adimensional, are able to draw conclusions about possible mechanisms that could select these parameters. If they are randomly drawn from a distribution by nature and if there are no fundamental scales in nature, we should be able to construct a distribution that doesn't refer to any particular scale. If we can't construct such a distribution, the parameters must be selected in some other way. As we shall see by considering QCD, the Coleman-Weinberg model and a totally asymptotically free theory, this constraint is non-trivial and requires us to find distributions that are RG invariant.

\section{RG invariant distributions}

We want to construct distributions for an adimensional theory's dimensionless parameters that don't refer to any particular dimensional scales. The dimensionless parameters of adimensional theories, however, run with
renormalization scale, $Q$. Naive choices of distribution would depend upon the choice of renormalization scale~\cite{Wells:2018sus}, as densities transform by a Jacobian rule. For example, the densities for a parameter $\alpha$ at scales $Q$ and $Q^\prime$ would be connected by
\begin{equation}\label{eq:jacobian_rule}
	p\left(\alpha(Q)\right) = p^\prime\left(\alpha(Q^\prime)\right) \, \left|\mathcal{J}\right|,
\end{equation}
where $\left|\mathcal{J}\right|$ is the Jacobian for the transformation between $\alpha(Q)$ and $\alpha(Q^\prime)$.\footnote{%
	This issue was raised by ref.~\cite{Wells:2018sus}. We don't agree with the proposed resolution, however, which was that probabilities behave as
	\begin{equation}
		p\left(\alpha(Q)\right) d\alpha(Q) = p^\prime\left(\alpha(Q^\prime)\right) d\alpha(Q^\prime),
	\end{equation}
	and that only probabilities, and not probability densities, are truly meaningful.
} %
In dimensional theories, particular scales could play a role in the distributions. For example, in a non-renormalizable effective theory valid at energies $E < M$ or in a grand unified theory (GUT) with unification scale $M$, there is a special scale and we could write distributions for parameters at that scale, $Q \approx M$.

In adimensional theories, we don't want the form of the distribution to depend on the scale $Q$, such that it doesn't have to be specified at any particular scale. We can achieve this by requiring that
\begin{equation}
	p = p^\prime,
\end{equation}
that is, that the two density functions are the same function. This is an example of an invariant distribution~(see e.g., ref.~\cite{10.1214/aoms/1177703583,Jaynes68priorprobabilities,doi:https://doi.org/10.1002/0471667196.ess1279.pub2,10.1214/18-BA1103}); other simple examples include
\begin{equation}
	p(x) \propto \frac1x,
\end{equation}
which is invariant under re-scalings $x \to A x$, and
\begin{equation}
	p(x) \propto \text{const.},
\end{equation}
which is invariant under shifts $x \to x + A$. Here we are seeking a distribution that is invariant under RG evolution. The scale and shift invariant distributions are examples of improper distributions: they cannot be normalized to one and we cannot sample from them.

Formally, invariant distributions are the right Haar measure of the transformation group~(see e.g., refs.~\cite{Easton1989,Berger1980,Robert2007}). This measure exists if the group is locally compact and leads to a proper distribution if and only if the group is compact. The number of distributions that can be found by a group invariance equals the size of the invariance group. 

For example, suppose that we considered dependent re-scalings for two parameters, $x \to A x$ and $y \to A y$. The size-one invariance group cannot uniquely dictate the prior for the two parameters. We obtain 
\begin{align}
	p(y \given x) &= \frac{f(x / y)}{y}\\
	p(x) &\propto \frac1x
\end{align}
for an arbitrary probability density function (PDF) $f$ because $x / y$ is invariant under the group. Similarly, if we were to consider dependent shifts, $x \to x + A$ and $y \to y + A$, we would obtain
\begin{align}
	p(y \given x) &= f(x - y)\\
	p(x) &\propto \text{const.}
\end{align}
for an arbitrary PDF $f$ because $x - y$ is invariant under the group. With this in mind, if there are RG invariant combinations of parameters, $\RGI$, the RG can only specify a distribution up to an arbitrary PDF of the invariant, $f(\RGI)$.

\section{QCD}

\begin{figure*}[t]
	\centering
	\begin{subfigure}[b]{0.49\textwidth}
		\centering
		\includegraphics[width=\textwidth]{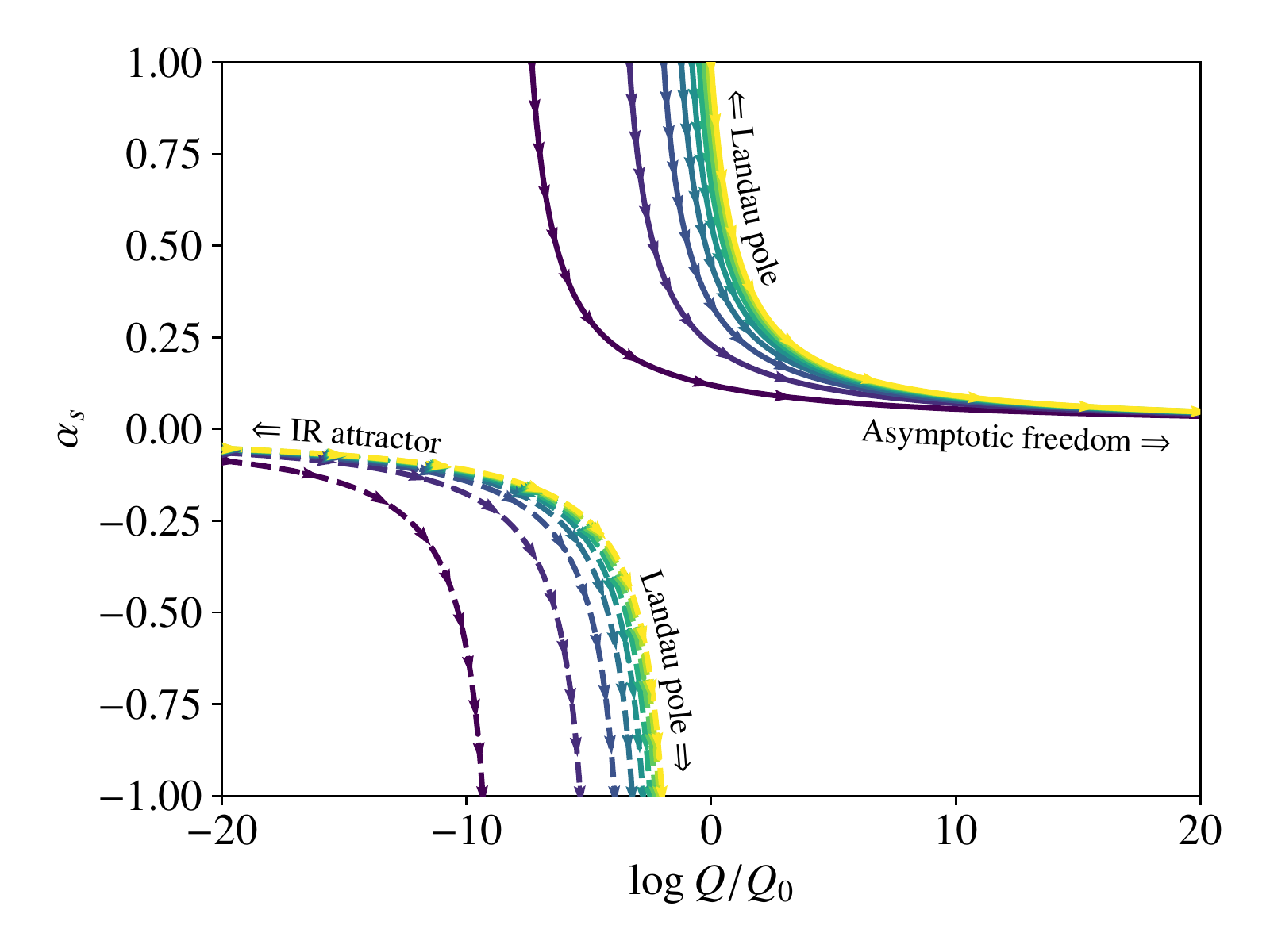}
		\caption{RG flow in QCD.}
		\label{fig:qcd}
	\end{subfigure}
	\hfill
	\begin{subfigure}[b]{0.49\textwidth}
		\centering
		\includegraphics[width=\textwidth]{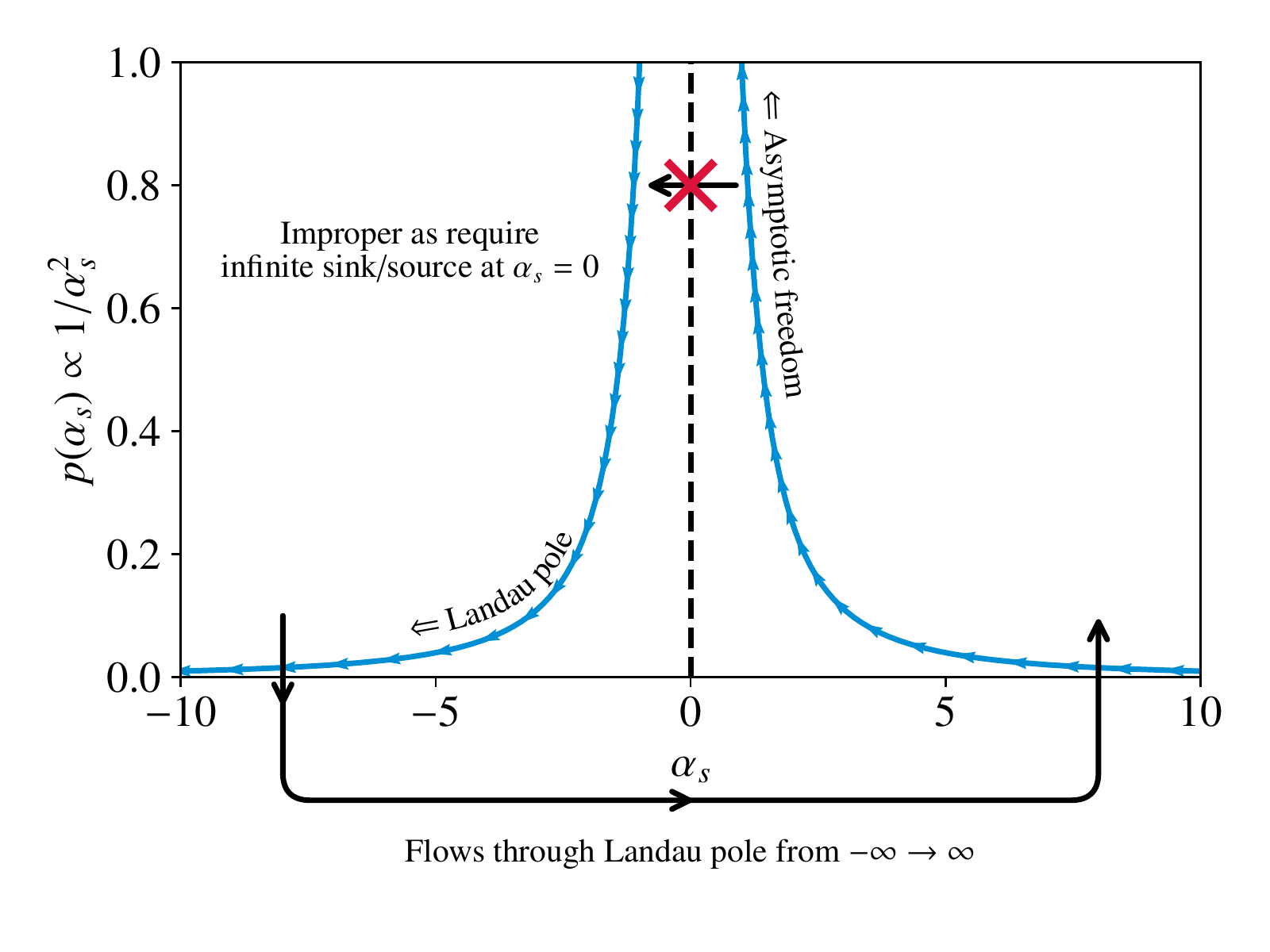}
		\caption{Invariant measure improper in QCD.}
		\label{fig:qcd_improper}
	\end{subfigure}
	\caption{The RG flow of the QCD gauge coupling leads to an improper invariant measure.}
\end{figure*}

Let's begin by considering quantum chromodynamics~(QCD), as in ref.~\cite{Wells:2018sus}. At one-loop, the RG equation for the strong coupling is
\begin{equation}\label{eq:qcd_beta}
	\dv{\alpha_s}{\ln Q} = - \beta_0 \alpha^2_s,
\end{equation}
where $ \beta_0 > 0$. Solving this differential equation yields the exact solution for $\alpha_s(Q)$ given $\alpha_s(Q^\prime)$,
\begin{equation}\label{eq:qcd_sol}
	\alpha_s(Q) = \frac{\alpha_s(Q^\prime)}{1 - \alpha_s(Q^\prime) \, \beta_0 \ln \left(Q^\prime / Q\right)}.
\end{equation}
We show this flow in \cref{fig:qcd}. This may be re-written as
\begin{equation}
	\alpha_s(Q) = \frac{1}{\beta_0 \ln \left(Q / \LQCD \right)},
\end{equation}
where we used the QCD scale,
\begin{equation}\label{eq:lambda_qcd}
	\LQCD = Q e^{- \frac{1}{\beta_0 \alpha_s(Q)} },
\end{equation}
which is a Landau pole in the RG evolution --- the coupling $\alpha_s$ cannot be evolved through the Landau pole at $Q = \LQCD$. This makes it hard to construct an RG invariant distribution for the coupling. We want the form of our distribution for $\alpha$ to be the same at any scale $Q$, but RG evolution to arbitrary $Q$ isn't possible. Formally, the transformations aren't closed and the Haar measure needn't exist.

We could consider Landau poles as fatal to the goal of constructing an RG invariant distribution, though Landau poles may be an artefact of the breakdown of perturbation theory. We could relax our requirement and seek RG invariant distributions only for $Q$ for which the RG evolution can be defined. Lastly, we could treat \cref{eq:qcd_sol} as a transformation valid for any $Q$, though this isn't a solution of \cref{eq:qcd_beta}.

For illustrative purposes, we pursue the latter approach. We may find the Jacobian from \cref{eq:qcd_sol}
\begin{equation}
	\jacobian = \left|\dv{\alpha_s(Q^\prime)}{\alpha_s(Q)}\right| = \left|\frac{\alpha_s(Q^\prime)}{\alpha_s(Q)}\right|^2.
\end{equation}
Through the Jacobian rule \cref{eq:jacobian_rule} this means that the probability densities for $\alpha_s$ at two different scales $Q$ and $Q^\prime$ are related by
\begin{align}
	p(\alpha_s(Q)) = p^\prime(\alpha_s(Q^\prime)) \, \jacobian.
\end{align}
Requiring that $p = p^\prime$ gives,
\begin{equation}\label{eq:qcd_invariant_distribution}
	p(\alpha_s) \propto \frac{1}{\alpha_s^2}.
\end{equation}
This is improper as the integral for the normalization diverges at $\alpha_s = 0$. We didn't include an arbitrary function of the QCD scale, $f(\LQCD)$ as we are only considering a measure on the fundamental parameter $\alpha_s$ and the transformation $\alpha_s(Q) \to \alpha_s(Q^\prime)$. The RG invariant $\LQCD$ would be relevant for a measure on~$(\alpha_s, Q)$ and the transformations $\alpha(Q) \to \alpha(Q^\prime)$ and $Q \to Q^\prime$. 

We can make sense of the fact that the distribution is improper by considering dimensional transmutation. The dimensionless parameter $\alpha_s$ may be traded for the QCD scale through \cref{eq:lambda_qcd}. Using the Jacobian rule and transforming from $\alpha_s$ to $\LQCD$, we find that \cref{eq:qcd_invariant_distribution} corresponds to
\begin{equation}\label{eq:p_lqcd}
	p(\LQCD) \propto \frac1{\LQCD}.
\end{equation}
This improper scale-invariant distribution was somewhat inevitable as there are no scales at all in our problem and thus on dimensional grounds \cref{eq:p_lqcd} was the only possibility. In \cref{fig:qcd_improper} we consider the flow of probability mass around the distribution \cref{eq:qcd_invariant_distribution}. Probability mass flows from the IR attractor at $-0$ to the UV attractor at $+0$ but cannot cross zero; hence, we require an infinite source and sink at $\alpha_s = 0$. Lastly, if we included $\alpha_s = 0$, a trivial invariant measure exists --- a Dirac mass at $\alpha_s = 0$.

In any case, the fact that the distribution was improper and the existence of the Landau pole are moot. Different choices of $\alpha_s(Q)$ describe an identical theory expressed in different units, as changing $\alpha_s(Q)$ may be compensated by scaling~$Q$. We may define a unit of energy to be $\LQCD$. This is a theory with no free parameters.

\section{Scalar QED}

We now consider the canonical example of radiative symmetry breaking: the Coleman-Weinberg model~\cite{PhysRevD.7.1888} of scalar quantum electrodynamics (QED). The model contains a complex scalar charged under a $U(1)$ gauge symmetry and an associated gauge boson described by the Lagrangian,
\begin{equation}
	\mathcal{L} = \left|D_\mu \phi\right|^2 - \frac{\lambda}{4!} |\phi|^4 - \frac14 F_{\mu\nu} F^{\mu\nu}.
\end{equation}
The tree-level scalar mass was set to zero, such that the model is adimensional. Masses are generated at one-loop by radiative symmetry breaking of the $U(1)$ gauge symmetry as logarithmic corrections to the scalar potential generate a non-trivial minimum.

\begin{figure*}[t]
	\centering
	\begin{subfigure}[b]{0.49\textwidth}
		\centering
		\includegraphics[width=\textwidth]{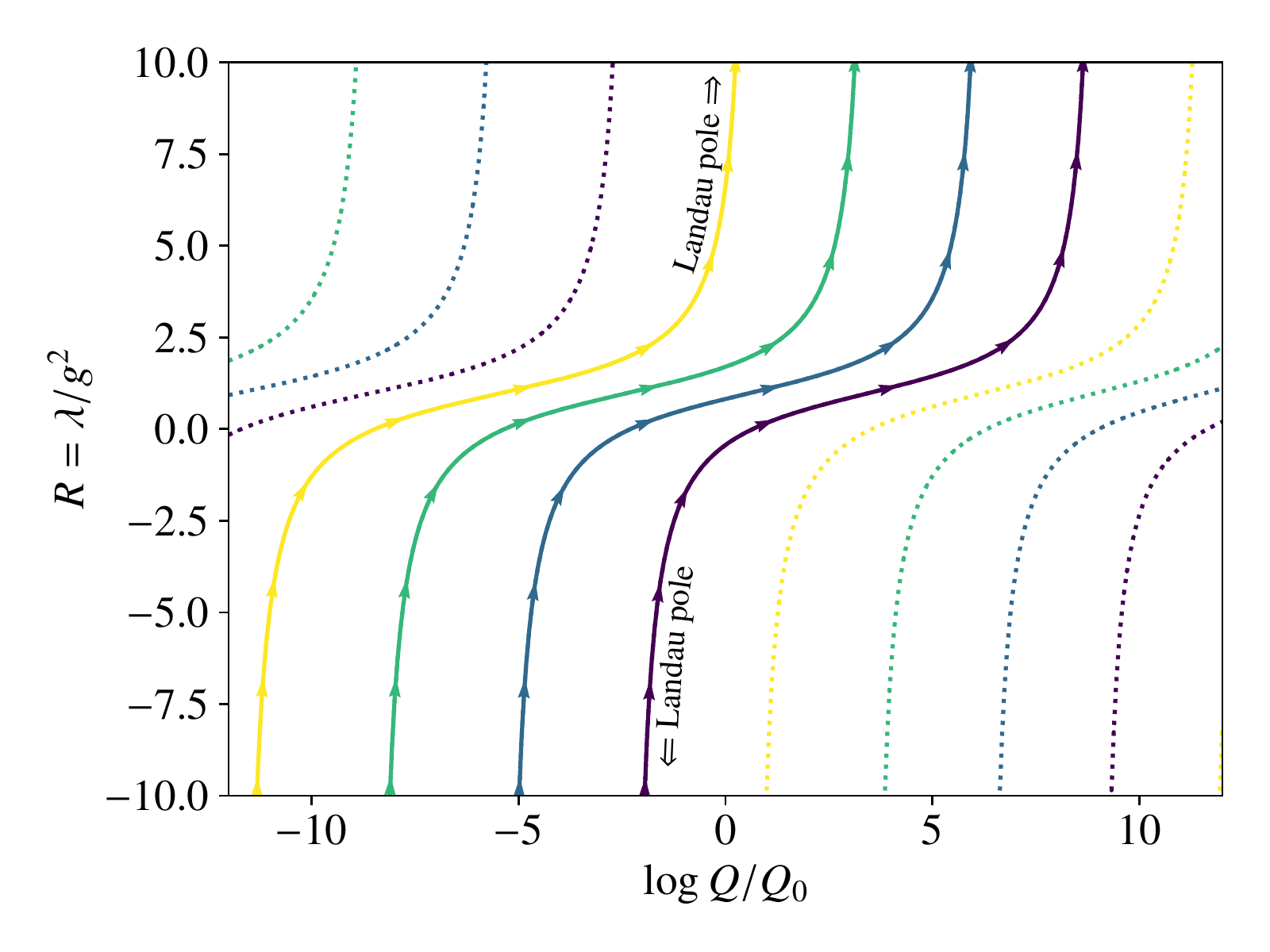}
		\caption{The Coleman-Weinberg model.}
		\label{fig:cw}
	\end{subfigure}
	\hfill
	\begin{subfigure}[b]{0.49\textwidth}
		\centering
		\includegraphics[width=\textwidth]{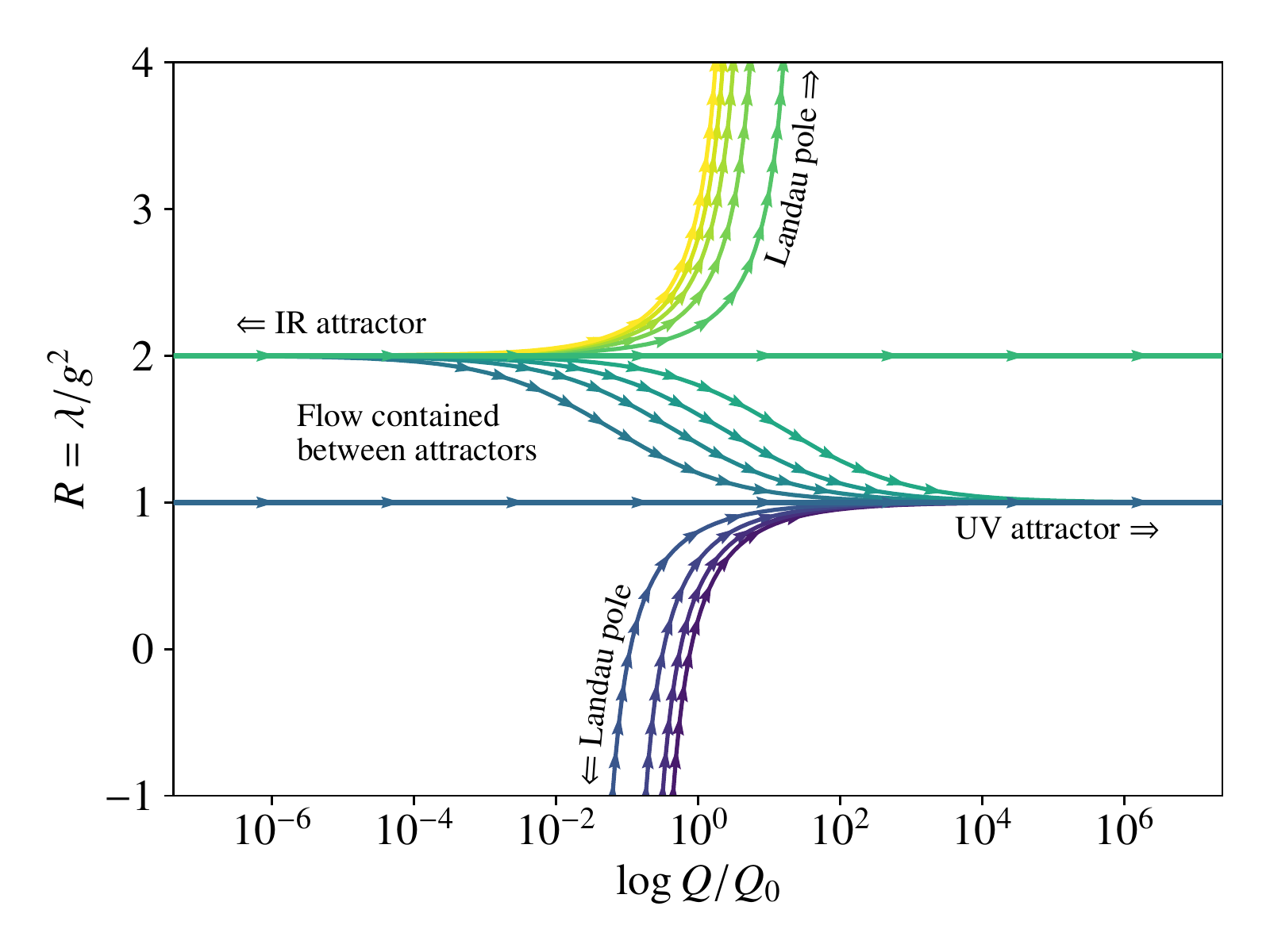}
		\caption{TAF theory with quartic and gauge couplings.}
		\label{fig:taf}
	\end{subfigure}
	\caption{RG flow with quartic and gauge coupling for the Coleman-Weinberg model and a TAF theory.}
\end{figure*}

The RG equations at one-loop are~\cite{Weinberg:1973am,PhysRevD.7.1888,Andreassen:2014eha}
\begin{align}
	\dv{\lambda}{\ln Q} & = \frac{10}{3 (4\pi)^2} \lambda^2 - \frac{12}{(4\pi)} \alpha \lambda + 36 \alpha^2, \label{eq:rge_lambda} \\
	\dv{\alpha}{\ln Q}  & = -\beta_0 \alpha^2, \label{eq:rge_alpha}
\end{align}
where $\beta_0 = - 2 / (12 \pi) $. The solutions are
\begin{align}\label{eq:sol_cw_alpha}
	\alpha(Q)  & = \frac{\alpha(Q^\prime)}{1 - \alpha(Q^\prime) \, \beta_0 \ln \left(Q^\prime / Q\right)},                         \\
	\lambda(Q) & = \gamma \alpha(Q) \tan\left[\frac{\sqrt{719}}{2}\ln \alpha(Q) + \Theta \right] + \delta \alpha(Q),\label{eq:sol_cw_lambda}
\end{align}
where the coefficients in \cref{eq:sol_cw_lambda} are
\begin{equation}
	\gamma = \frac{\sqrt{719}}{10} 4\pi \quad\text{and}\quad \delta = \frac{19}{10} 4\pi .
\end{equation}
The quantity $\Theta$ may be chosen to fix $\lambda(Q^\prime)$ through
\begin{equation}\label{eq:rgi_theta}
	\Theta = \arctan\left[
		\frac{\lambda(Q^\prime) - \delta \alpha(Q^\prime)}{\gamma \alpha(Q^\prime)}
		\right]
	- \frac{\sqrt{719}}{2} \ln \alpha(Q^\prime).
\end{equation}
This parameter is in fact an RG invariant since it may be determined from \cref{eq:sol_cw_lambda} at any scale. The tangent in \cref{eq:sol_cw_lambda} means that $\lambda$ swings rapidly from $-\infty$ to $\infty$ between Landau poles. Thus scalar QED describes a narrow energy range between two Landau poles. This is shown in \cref{fig:cw}.

As before, for illustrative purposes we continue regardless of the Landau poles. Treating the transformations \cref{eq:sol_cw_alpha,eq:sol_cw_lambda} as valid for any $Q$, we compute the associated Jacobian,
\begin{align}
	\jacobian & = \left|\dv{(\alpha(Q^\prime), \lambda(Q^\prime))}{(\alpha(Q), \lambda(Q))}\right| \\
	          & =\frac{
		\alpha(Q^\prime)^3 \left[1 + \left(\frac{\lambda(Q^\prime) - \delta \alpha(Q^\prime)}{\gamma \alpha(Q^\prime)}\right)^2\right]
	}
	{
		\alpha(Q)^3 \left[1 + \left(\frac{\lambda(Q) - \delta \alpha(Q)}{\gamma \alpha(Q)}\right)^2\right]
	}
\end{align}
The distributions are related by
\begin{align}
	p(\lambda(Q), \alpha(Q)) = p^\prime(\lambda(Q^\prime), \alpha(Q^\prime)) \, \jacobian
\end{align}
We may construct an RG invariant distribution for the couplings by requiring that $p$ and $p^\prime$ are the same function,
\begin{equation}
	p(\lambda, \alpha) \propto \frac{f(\Theta)}{
		\alpha^3 \left[1 + \left(\frac{\lambda - \delta \alpha}{\gamma \alpha}\right)^2\right]
	}
\end{equation}
where $f$ is an arbitrary PDF of the RG invariant combination of $\lambda$ and $\alpha$. We may marginalize $\lambda$, finding,
\begin{align}
	p(\lambda \given \alpha) & = \frac{f(\Theta)}{\gamma \alpha \pi\left[1 + \left(\frac{\lambda - \delta \alpha}{\gamma \alpha}\right)^2\right]},\label{eq:p_l} \\
	p(\alpha)                & \propto \frac{1}{\alpha^2}, \label{eq:p_g}
\end{align}
We recognize \cref{eq:p_l} as a Cauchy distribution located at $\delta \alpha$ with scale parameter $\gamma \alpha$ that is multiplied by an arbitrary PDF of the RG invariant. 

As for QCD, the distribution for the gauge coupling in \cref{eq:p_g} is improper. We could try to we side-step this by using the gauge coupling to fix our units. In analogy with QCD, by dimensional transmutation we may change variable,
\begin{equation}\label{eq:lambda_cw}
	\Lambda = Q e^{\frac{1}{\beta_0 \alpha^2}},
\end{equation}
and fix $\Lambda$ arbitrarily as it only defines a choice of units. We are left with a proper RG-invariant distribution, \cref{eq:p_l}, for the quartic coupling $\lambda$. Practically, this only partially helps us make predictions from scalar QED. Knowledge of $\alpha(Q)$ and $\lambda(Q)$ at an arbitrary $Q$ isn't enough; we need to know the validity of the theory, e.g., the scale of a Landau pole in the tangent defining a limit of the theory's validity.

\section{Totally asymptotically free theory}\label{sec:taf}

For our purposes, QCD was trivial and scalar QED suffered from Landau poles in the IR and UV such that transformations couldn't form a closed group. We thus now turn to a similar theory with no Landau poles in the UV. The easiest theories of this type to construct and analyse are totally asymptotically free theories~\cite{Giudice:2014tma}. To make a simple model, we consider a scalar with a non-Abelian gauge interaction.

The RG equation for the gauge coupling is identical to that in \cref{eq:rge_alpha}, though we consider $\beta_0 > 0$. We write an agnostic RG equation for the quartic,
\begin{equation}
	\dv{\lambda}{\ln Q} = s_\lambda \lambda^2 - s_{\lambda g} \lambda g^2 + s_g g^4, \label{eq:rge_lambda_general}
\end{equation}
where the coefficients $s$ must be positive for any QFT. This generalizes \cref{eq:rge_lambda} and follows ref.~\cite{Giudice:2014tma}. The coefficients can be parameterized by
\begin{align}
	C & \equiv \frac{s_g}{\beta_0},                   \\
	D & \equiv \frac{s_{\lambda g} - \beta_0}{2 s_g}, \\
	E & \equiv D^2 - \frac{s_\lambda}{s_g}.
\end{align}
For $E < 0$, we obtain a general solution of the form in \cref{eq:sol_cw_lambda}.
For $E=0$, we obtain trivial fixed-flow behaviour. Lastly, for $E > 0$, the solution may be written as
\begin{align}
	R(Q)      & = R_\text{IR} + \left( R_\text{UV} - R_\text{IR}\right) F(\alpha),                                             \\
	F(\alpha) & \equiv \frac12\left[ 1 - \tanh\left(C \sqrt{E} \ln \alpha(Q) + \Theta \right)\right], \label{eq:sol_pos_lambda}
\end{align}
where $R(Q) \equiv {\lambda(Q)}/{(4\pi\alpha(Q))}$ and
\begin{equation}
	R_\text{IR} = \frac{1}{D - \sqrt{E}} \quad\text{and}\quad R_\text{UV} = \frac{1}{D + \sqrt{E}}.
\end{equation}
We pursue this $E > 0$ case to avoid Landau poles. The function $F(\alpha)$ flows from $0$ in the IR to $1$ in the UV such that, as shown in \cref{fig:taf}, we encounter IR and UV attractors at $R_\text{IR/UV}$. If at any scale the ratio lies inside $[R_\text{UV}, R_\text{IR}]$, it stays trapped inside that interval and there are no Landau poles. As before, $\Theta$ is an RG invariant,
\begin{equation}
	\Theta = \text{arctanh}\left[1 - 2\left( \frac{R(Q^\prime) - R_\text{IR}}{R_\text{UV}- R_\text{IR}}\right)\right] - C\sqrt{E}\log\alpha(Q^\prime)
\end{equation}
that controls $R(Q^\prime)$.

After computing the Jacobian, we find an RG invariant measure on $(R_\text{UV}, R_\text{IR})$,
\begin{align}
	p(R \given \alpha) & = \frac12 \frac{ ( R_\text{IR} - R_\text{UV})}{( R_\text{IR} - R) \, (R - R_\text{UV})} \, f(\Theta) ,\label{eq:rgi_taf} \\
	p(\alpha)          & \propto \frac{1}{\alpha^2},
\end{align}
for an arbitrary PDF $f(\Theta)$. If we were to consider $[R_\text{UV}, R_\text{IR}]$, Dirac masses at the fixed points would be possible as well, $\delta(R - R_\text{IR/UV})$. Although the PDF $f(\Theta)$ is arbitrary, the fact it can only be a function of $\Theta$ is restrictive and analogous to physics problems with spherical symmetry in which a result must depend only on the radius. 

The attractors led to poles in the RG invariant measure in \cref{eq:rgi_taf}. Whilst this behaviour is intuitive, it risks making the distribution improper. However, $f(\Theta)$ is a choice of measure on the RG invariant and must be proper,
\begin{equation}
	\int_{-\infty}^\infty f(\Theta) \,d\Theta = 1.\label{eq:f_proper}
\end{equation}
 That means it must decay to zero faster than $1 / \Theta$. Making an expansion near the poles,
\begin{equation}
	\Theta \propto \log(R - R_\text{IR/UV})
\end{equation}
and so $p(R\given \alpha)$ must grow slower around the poles than,
\begin{equation}
	\frac1{\log(R - R_\text{IR/UV})} \frac{1}{(R_\text{IR} - R) \, (R - R_\text{UV})}.
\end{equation}
and so the invariant measure is proper, providing that $f(\Theta)$ was proper. 

\begin{figure}[t]
	\centering
	\includegraphics[width=0.8\linewidth]{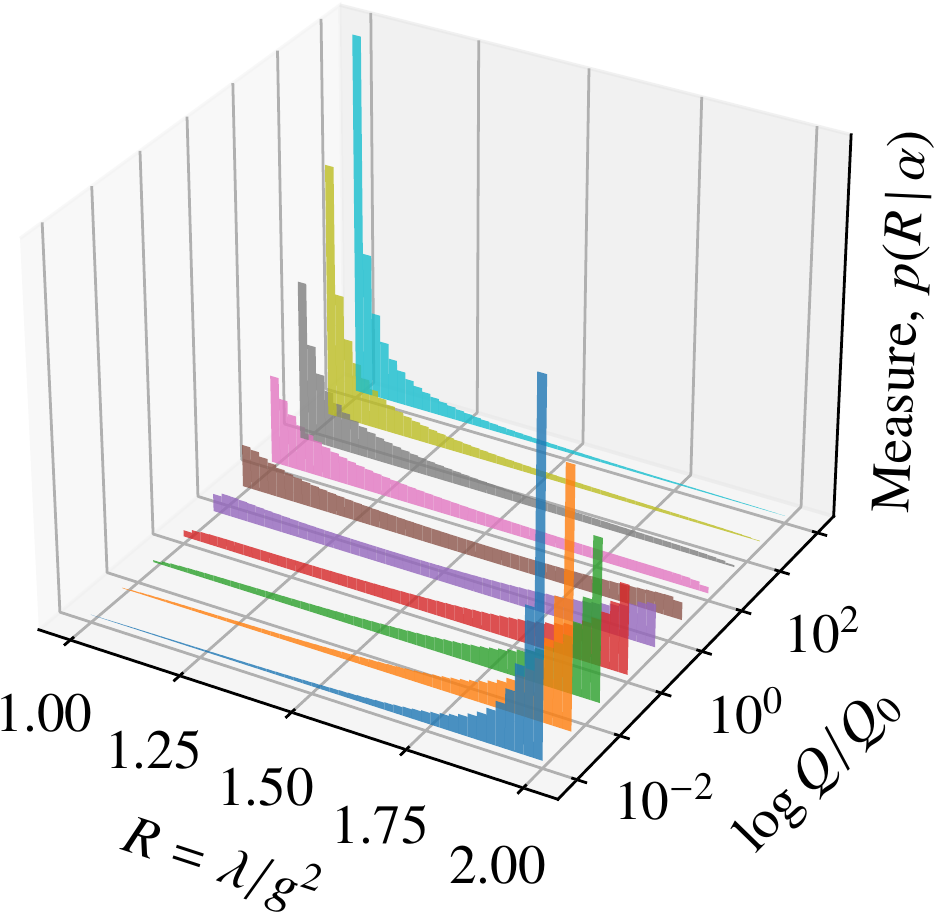}
	\caption{Flow of RG invariant measure between IR and UV attractors.}
	\label{fig:taf_measure}
\end{figure}

The form of conditional distribution \cref{eq:rgi_taf} doesn't change under RG evolution, though the shape of the distribution flows. This happens because $\Theta(R, \alpha)$ for fixed $R$ changes as $\alpha$ flows. To achieve the `wrong' value of $R$ --- $R \simeq R_\text{UV}$ in the IR or $R \simeq R_\text{IR}$ in the UV --- requires $|\Theta| \to \infty$. By \cref{eq:f_proper}, $\lim_{\Theta \to \infty} f(\Theta) = 0$, such that these `wrong' values are always suppressed by $f(\Theta)$. Thus, generically the factor $f(\Theta)$ pushes probability from $R_\text{IR}$ to $R_\text{UV}$ as $\ln\alpha$ flows from $\infty$ to $-\infty$. We show an example in \cref{fig:taf_measure} for $f(\Theta) = \mathcal{N}(0, 1)$, a standard normal distribution.

This analysis was at one-loop. The principles are the same at any order and we don't expect further corrections to spoil the fact that \cref{eq:rgi_taf} was proper, spoil the existence of an RG invariant, or prevent us from using $\alpha$ to define a system of units. We anticipate, however, that at higher order we would require special techniques to obtain RG invariant distributions, as it would be challenging to compute the Jacobian and identify the RG invariants. Approximate and numerical results, rather than exact analytic results, may be necessary. 

\section{Conclusions}\label{sec:conclusions}

We considered the origins of fundamental parameters in adimensional models. We argued that if the parameters originate as random draws from a distribution, that distribution must be RG invariant to avoid introducing massive scales. If an invariant measure doesn't exist for a model and that model is truly adimensional, the model's parameters cannot originate as random draws. On the other hand, if invariant measures exist for a model and we can compute the form of the measure, the model's parameters must originate as draws from this form of measure, if they originate as random draws. 

We continued by attempting to construct RG invariant measures for three concrete theories: QCD, the Coleman-Weinberg model, and a totally asymptotically free theory. We encountered two problems. First, to be able to sample from the invariant distributions, they should be proper, but the distribution for the anomalously generated scale in QCD was improper. This isn't necessarily an issue, though, as a single anomalously generated scale only determines a system of units. Second, the presence of Landau poles makes it impossible to construct RG invariant distributions for parameters, since the parameters themselves aren't defined at scales beyond the Landau pole and RG transformations can't form a closed group. This meant that we couldn't construct a satisfactory RG invariant distribution in the Coleman-Weinberg model. 

The absence of Landau poles in the totally asymptotically free theory, however, allowed us to construct an invariant measure for the quartic coupling. The measure was an arbitrary choice of PDF for an RG invariant combination of the quartic and gauge coupling combined with poles at the IR and UV fixed points in the RG flow of the quartic. Thus this quartic could originate as a random draw and if it does, we know the form of distribution that it must originate from. The requirement that a model is adimensional is so strong that it opens a window on the origin of the model's parameters, something usually beyond us.

\acknowledgments

I would like to thank Luca Marzola for discussions and feedback.

\bibliographystyle{JHEP}
\bibliography{refs}

\providecommand{\href}[2]{#2}\begingroup\raggedright\begin{thebibliography}{10}

\bibitem{Foot:2007iy}
R.~Foot, A.~Kobakhidze, K.L.~McDonald and R.R.~Volkas, \emph{{A Solution to the
  hierarchy problem from an almost decoupled hidden sector within a classically
  scale invariant theory}},
  \href{https://doi.org/10.1103/PhysRevD.77.035006}{\emph{Phys. Rev. D}
  {\bfseries 77} (2008) 035006}
  [\href{https://arxiv.org/abs/0709.2750}{{\ttfamily 0709.2750}}].

\bibitem{Heikinheimo:2013fta}
M.~Heikinheimo, A.~Racioppi, M.~Raidal, C.~Spethmann and K.~Tuominen,
  \emph{{Physical Naturalness and Dynamical Breaking of Classical Scale
  Invariance}}, \href{https://doi.org/10.1142/S0217732314500771}{\emph{Mod.
  Phys. Lett. A} {\bfseries 29} (2014) 1450077}
  [\href{https://arxiv.org/abs/1304.7006}{{\ttfamily 1304.7006}}].

\bibitem{Gabrielli:2013hma}
E.~Gabrielli, M.~Heikinheimo, K.~Kannike, A.~Racioppi, M.~Raidal and
  C.~Spethmann, \emph{{Towards Completing the Standard Model: Vacuum Stability,
  EWSB and Dark Matter}},
  \href{https://doi.org/10.1103/PhysRevD.89.015017}{\emph{Phys. Rev. D}
  {\bfseries 89} (2014) 015017}
  [\href{https://arxiv.org/abs/1309.6632}{{\ttfamily 1309.6632}}].

\bibitem{Englert:2013gz}
C.~Englert, J.~Jaeckel, V.V.~Khoze and M.~Spannowsky, \emph{{Emergence of the
  Electroweak Scale through the Higgs Portal}},
  \href{https://doi.org/10.1007/JHEP04(2013)060}{\emph{JHEP} {\bfseries 04}
  (2013) 060} [\href{https://arxiv.org/abs/1301.4224}{{\ttfamily 1301.4224}}].

\bibitem{Kannike:2014mia}
K.~Kannike, A.~Racioppi and M.~Raidal, \emph{{Embedding inflation into the
  Standard Model - more evidence for classical scale invariance}},
  \href{https://doi.org/10.1007/JHEP06(2014)154}{\emph{JHEP} {\bfseries 06}
  (2014) 154} [\href{https://arxiv.org/abs/1405.3987}{{\ttfamily 1405.3987}}].

\bibitem{Salvio:2020axm}
A.~Salvio, \emph{{Dimensional Transmutation in Gravity and Cosmology}},
  \href{https://doi.org/10.1142/S0217751X21300064}{\emph{Int. J. Mod. Phys. A}
  {\bfseries 36} (2021) 2130006}
  [\href{https://arxiv.org/abs/2012.11608}{{\ttfamily 2012.11608}}].

\bibitem{Wilson:1971dc}
K.G.~Wilson and M.E.~Fisher, \emph{{Critical exponents in 3.99 dimensions}},
  \href{https://doi.org/10.1103/PhysRevLett.28.240}{\emph{Phys. Rev. Lett.}
  {\bfseries 28} (1972) 240}.

\bibitem{Wilson:1971vs}
K.G.~Wilson, \emph{{Feynman graph expansion for critical exponents}},
  \href{https://doi.org/10.1103/PhysRevLett.28.548}{\emph{Phys. Rev. Lett.}
  {\bfseries 28} (1972) 548}.

\bibitem{Wilson:1973jj}
K.G.~Wilson and J.B.~Kogut, \emph{{The renormalization group and the $\epsilon$
  expansion}}, \href{https://doi.org/10.1016/0370-1573(74)90023-4}{\emph{Phys.
  Rept.} {\bfseries 12} (1974) 75}.

\bibitem{Wilson:1974mb}
K.G.~Wilson, \emph{{The Renormalization Group: Critical Phenomena and the Kondo
  Problem}}, \href{https://doi.org/10.1103/RevModPhys.47.773}{\emph{Rev. Mod.
  Phys.} {\bfseries 47} (1975) 773}.

\bibitem{Carvalho:2022hca}
P.R.S.~Carvalho, \emph{{Experimental validation of nonextensive statistical
  field theory: Applications to manganites}},
  \href{https://doi.org/10.1016/j.physletb.2023.137683}{\emph{Phys. Lett. B}
  {\bfseries 838} (2023) 137683}
  [\href{https://arxiv.org/abs/2211.07577}{{\ttfamily 2211.07577}}].

\bibitem{Bardeen:1995kv}
W.A.~Bardeen, \emph{{On naturalness in the Standard Model}},  in \emph{{Ontake
  Summer Institute on Particle Physics}}, 8, 1995,
  \href{https://inspirehep.net/files/3be9a25854a9bf89b5ad70715029ca5c}{URL}.

\bibitem{Salvio:2014soa}
A.~Salvio and A.~Strumia, \emph{{Agravity}},
  \href{https://doi.org/10.1007/JHEP06(2014)080}{\emph{JHEP} {\bfseries 06}
  (2014) 080} [\href{https://arxiv.org/abs/1403.4226}{{\ttfamily 1403.4226}}].

\bibitem{Salvio:2017qkx}
A.~Salvio and A.~Strumia, \emph{{Agravity up to infinite energy}},
  \href{https://doi.org/10.1140/epjc/s10052-018-5588-4}{\emph{Eur. Phys. J.}
  {\bfseries C78} (2018) 124}
  [\href{https://arxiv.org/abs/1705.03896}{{\ttfamily 1705.03896}}].

\bibitem{Salvio:2018crh}
A.~Salvio, \emph{{Quadratic Gravity}},
  \href{https://doi.org/10.3389/fphy.2018.00077}{\emph{Front. in Phys.}
  {\bfseries 6} (2018) 77} [\href{https://arxiv.org/abs/1804.09944}{{\ttfamily
  1804.09944}}].

\bibitem{Donoghue:2021cza}
J.F.~Donoghue and G.~Menezes, \emph{{On quadratic gravity}},
  \href{https://doi.org/10.1393/ncc/i2022-22026-7}{\emph{Nuovo Cim. C}
  {\bfseries 45} (2022) 26} [\href{https://arxiv.org/abs/2112.01974}{{\ttfamily
  2112.01974}}].

\bibitem{Kannike:2015apa}
K.~Kannike, G.~H\"utsi, L.~Pizza, A.~Racioppi, M.~Raidal, A.~Salvio et~al.,
  \emph{{Dynamically Induced Planck Scale and Inflation}},
  \href{https://doi.org/10.1007/JHEP05(2015)065}{\emph{JHEP} {\bfseries 05}
  (2015) 065} [\href{https://arxiv.org/abs/1502.01334}{{\ttfamily
  1502.01334}}].

\bibitem{Salvio:2015gsi}
A.~Salvio and A.~Strumia, \emph{{Quantum mechanics of 4-derivative theories}},
  \href{https://doi.org/10.1140/epjc/s10052-016-4079-8}{\emph{Eur. Phys. J. C}
  {\bfseries 76} (2016) 227}
  [\href{https://arxiv.org/abs/1512.01237}{{\ttfamily 1512.01237}}].

\bibitem{Raidal:2016wop}
M.~Raidal and H.~Veerm\"ae, \emph{{On the Quantisation of Complex Higher
  Derivative Theories and Avoiding the Ostrogradsky Ghost}},
  \href{https://doi.org/10.1016/j.nuclphysb.2017.01.024}{\emph{Nucl. Phys. B}
  {\bfseries 916} (2017) 607}
  [\href{https://arxiv.org/abs/1611.03498}{{\ttfamily 1611.03498}}].

\bibitem{Strumia:2017dvt}
A.~Strumia, \emph{{Interpretation of quantum mechanics with indefinite norm}},
  \href{https://doi.org/10.3390/physics1010003}{\emph{MDPI Physics} {\bfseries
  1} (2019) 17} [\href{https://arxiv.org/abs/1709.04925}{{\ttfamily
  1709.04925}}].

\bibitem{Gross:2020tph}
C.~Gross, A.~Strumia, D.~Teresi and M.~Zirilli, \emph{{Is negative kinetic
  energy metastable?}},
  \href{https://doi.org/10.1103/PhysRevD.103.115025}{\emph{Phys. Rev. D}
  {\bfseries 103} (2021) 115025}
  [\href{https://arxiv.org/abs/2007.05541}{{\ttfamily 2007.05541}}].

\bibitem{Donoghue:2021eto}
J.F.~Donoghue and G.~Menezes, \emph{{Ostrogradsky instability can be overcome
  by quantum physics}},
  \href{https://doi.org/10.1103/PhysRevD.104.045010}{\emph{Phys. Rev. D}
  {\bfseries 104} (2021) 045010}
  [\href{https://arxiv.org/abs/2105.00898}{{\ttfamily 2105.00898}}].

\bibitem{Wells:2018sus}
J.D.~Wells, \emph{{Naturalness, Extra-Empirical Theory Assessments, and the
  Implications of Skepticism}},
  \href{https://doi.org/10.1007/s10701-018-0220-x}{\emph{Found. Phys.}
  {\bfseries 49} (2019) 991}
  [\href{https://arxiv.org/abs/1806.07289}{{\ttfamily 1806.07289}}].

\bibitem{10.1214/aoms/1177703583}
J.~Hartigan, \emph{{Invariant Prior Distributions}},
  \href{https://doi.org/10.1214/aoms/1177703583}{\emph{The Annals of
  Mathematical Statistics} {\bfseries 35} (1964) 836 }.

\bibitem{Jaynes68priorprobabilities}
E.T.~Jaynes, \emph{Prior probabilities}, {\emph{IEEE Transactions on Systems
  Science and Cybernetics} {\bfseries 4} (1968) 227,
  \href{https://bayes.wustl.edu/etj/articles/prior.pdf}{URL}}.

\bibitem{doi:https://doi.org/10.1002/0471667196.ess1279.pub2}
A.P.~Dawid, \emph{Invariant prior distributions},  in \emph{Encyclopedia of
  Statistical Sciences}, Wiley (2006),
  \href{https://doi.org/10.1002/0471667196.ess1279.pub2}{DOI}.

\bibitem{10.1214/18-BA1103}
G.~Consonni, D.~Fouskakis, B.~Liseo and I.~Ntzoufras, \emph{{Prior
  Distributions for Objective Bayesian Analysis}},
  \href{https://doi.org/10.1214/18-BA1103}{\emph{Bayesian Analysis} {\bfseries
  13} (2018) 627 }.

\bibitem{Easton1989}
M.L.~Easton, \emph{Group invariance in applications in statistics}, Institute
  of Mathematical Statistics and American Statistical Association (1989),
  \href{https://doi.org/10.1214/cbms/1462061029}{DOI}.

\bibitem{Berger1980}
J.O.~Berger, \emph{Invariance},  in \emph{Statistical Decision Theory},
  pp.~237--280, Springer New York (1980),
  \href{https://doi.org/10.1007/978-1-4757-1727-3_6}{DOI}.

\bibitem{Robert2007}
C.P.~Robert, \emph{Invariance, haar measures, and equivariant estimators},  in
  \emph{The Bayesian Choice: From Decision-Theoretic Foundations to
  Computational Implementation}, (New York, NY), pp.~427--455, Springer New
  York (2007), \href{https://doi.org/10.1007/0-387-71599-1_9}{DOI}.

\bibitem{PhysRevD.7.1888}
S.~Coleman and E.J.~Weinberg, \emph{Radiative corrections as the origin of
  spontaneous symmetry breaking},
  \href{https://doi.org/10.1103/PhysRevD.7.1888}{\emph{Phys. Rev. D} {\bfseries
  7} (1973) 1888}.

\bibitem{Weinberg:1973am}
E.J.~Weinberg, \emph{{Radiative corrections as the origin of spontaneous
  symmetry breaking}}, Ph.D. thesis, Harvard U., 1973.
\newblock \href{https://arxiv.org/abs/hep-th/0507214}{{\ttfamily
  hep-th/0507214}}.

\bibitem{Andreassen:2014eha}
A.~Andreassen, W.~Frost and M.D.~Schwartz, \emph{{Consistent Use of Effective
  Potentials}}, \href{https://doi.org/10.1103/PhysRevD.91.016009}{\emph{Phys.
  Rev. D} {\bfseries 91} (2015) 016009}
  [\href{https://arxiv.org/abs/1408.0287}{{\ttfamily 1408.0287}}].

\bibitem{Giudice:2014tma}
G.F.~Giudice, G.~Isidori, A.~Salvio and A.~Strumia, \emph{{Softened Gravity and
  the Extension of the Standard Model up to Infinite Energy}},
  \href{https://doi.org/10.1007/JHEP02(2015)137}{\emph{JHEP} {\bfseries 02}
  (2015) 137} [\href{https://arxiv.org/abs/1412.2769}{{\ttfamily 1412.2769}}].

\end{thebibliography}\endgroup

\end{document}